\documentclass[aps,pra,reprint,superscriptaddress,longbibliography,floatfix]{revtex4-2}
\usepackage[T1]{fontenc}
\usepackage[utf8]{inputenc}
\usepackage{lmodern}
\usepackage{amsmath,amssymb,bm}
\usepackage{graphicx,xcolor}
\usepackage[colorlinks=true,linkcolor=blue!55!black,citecolor=blue!55!black,urlcolor=blue!55!black]{hyperref}

\newcommand{\fs}{f_{\mathrm{s}}}
\newcommand{\Te}{T_{\mathrm{e}}}

\begin{document}

\title{High-speed laser speckle contrast imaging of the human retina}

\author{Zacharie Auray}
\affiliation{Institut Langevin, CNRS, ESPCI Paris, PSL University, Paris, France}
\author{Jules Guillou}
\affiliation{EPITA Research Laboratory, Le Kremlin-Bic\^{e}tre, France}
\author{Basile Malbois}
\affiliation{Institut Langevin, CNRS, ESPCI Paris, PSL University, Paris, France}
\author{Maxime Boy-Arnould}
\affiliation{Institut Langevin, CNRS, ESPCI Paris, PSL University, Paris, France}
\author{Michael Atlan}
\affiliation{Institut Langevin, CNRS, ESPCI Paris, PSL University, Paris, France}
\affiliation{EPITA Research Laboratory, Le Kremlin-Bic\^{e}tre, France}

\date{\today}

\begin{abstract}
We demonstrate real-time laser speckle contrast imaging of the human retina
using a Phantom S980 streaming camera at $3840\times2160$ pixels,
8-bit depth, and 1000 frames per second, with a $988\,\mu\mathrm{s}$
exposure. We report four recordings of the peripapillary retina around
the optic disc in the right eye of one healthy 25-year-old emmetropic man,
acquired without pharmacological pupil dilation,
using 852-nm diffuse Maxwellian illumination at an entering optical
power of 5.0 mW. Each recording contained 4096 frames (4.10 s), and all
16384 frames in the reported recordings were retained. Open-source Holoflow software on
an NVIDIA RTX 5090 computes $K=\sigma/\mu$ over $5\times5$ spatial
pixels and nonoverlapping blocks of 16 frames, producing 62.5 contrast
maps/s with 60-frame/s rendering and no additional averaging. The nominal
raw-input throughput is 8.29 Gpixel/s. Main retinal arterial and venous
branches and smaller peripapillary vessels were visualized. Variations
during the cardiac cycle were visible in the principal retinal arteries
and veins, as illustrated in the Supplementary Videos. Regional
inverse-contrast traces show an arterial dicrotic notch and spontaneous
venous pulsation. This
single-volunteer demonstration establishes an operating point for
high-throughput retinal speckle-contrast visualization. Physiological
observations are descriptive; contrast is an instrument-dependent
observable rather than a calibrated blood-flow velocity.

\end{abstract}

\maketitle

\section{Introduction}

Blood-flow dynamics provide functional information about the retinal
circulation that complements vascular morphology. A useful measurement must
capture temporal variations while preserving enough spatial information to
distinguish vascular regions. Laser speckle contrast imaging (LSCI) addresses
this problem through the statistics of coherent-light intensity patterns.
Early work demonstrated flow visualization from single-exposure speckle
photography, including early human fundus imaging, followed by digital
laser speckle contrast analysis
(LASCA)~\cite{Fercher1981,Briers1982,Briers1996}. Retinal implementations subsequently
used temporal statistics and endoscopic imaging to study blood-flow-related
signals and physiological responses
\cite{Cheng2007,Cheng2008,Srienc2010,Ponticorvo2013}.
Kilohertz camera acquisition has also been used in nonretinal multi-exposure
speckle imaging, with a distinct rate for the processed output
\cite{Hultman2018,Hultman2020Realtime}. Efficient rolling-statistics
algorithms, graphics-processor implementations, and dedicated programmable
hardware are established precedents for real-time speckle processing
\cite{Tom2008Processing,Liu2008GPU,Yang2011GPU,Jiang2011FPGA}.

The term LSCI describes the imaging modality, whereas LASCA emphasizes the
contrast calculation. Spatial, temporal, and pooled spatiotemporal estimators
use different ensembles of intensity samples. Spatiotemporal contrast
estimation combines neighboring pixels and successive frames within one
statistical neighborhood~\cite{Qiu2010}. Its precision depends on the number
and correlation of the sampled speckles, the detector sampling, and the
numerical implementation~\cite{Duncan2008Statistics,Kirkpatrick2008,Thompson2011,Chammas2025}.
The number of samples in a kernel is consequently not equivalent to the
number of independent observations.

Human retinal speckle imaging already supports pulsatile measurements. A
clinical laser speckle flowgraphy (LSFG) implementation acquires 118 images
over approximately four seconds at 30 frames/s with a 2-ms camera exposure,
and reports mean blur rate as a relative blood-flow index~\cite{Luft2016}.
Handheld retinal LSCI has also been investigated~\cite{Rege2018}. The XyCAM
RI implementation acquired human retinal images at 82 frames/s and estimated
contrast using a $5\times5$ spatial neighborhood spanning 15 frames, enabling
arterial and venous waveform analysis~\cite{Cho2020}. Multi-exposure human
retinal imaging has instead combined several camera exposure times to
estimate decorrelation-related quantities and assess repeatability
\cite{Qiu2022}.

The numerical outputs of these approaches are related but distinct.
LSFG reports mean blur rate (MBR) and its regional or pulse-wave summaries,
whereas relative flow volume (RFV) integrates a background-corrected
vessel profile~\cite{Sugiyama2014,Shiga2014}. Validation against laser
Doppler velocimetry and Doppler optical coherence tomography supports
RFV as a relative retinal flow index, with a response that depends on the
operating range~\cite{Shiga2014,Luft2016Validation}. Retinal and optic
nerve head autoregulatory responses have also been studied using
LSFG~\cite{Witkowska2017}. Numerical quantification and repeatability
should therefore be distinguished from calibration in absolute velocity or
volumetric-flow units.

Preclinical studies have established additional applications of retinal
LSCI, including pharmacological vascular responses and vasomotion
\cite{Neganova2016,Neganova2017}, and experimental validation of relative
flow measurements and wide-field instrumentation
\cite{Patel2020,Patel2021}. More recently, Tate and colleagues
reported murine retinal recordings at 376.2 frames/s with 2.5-ms exposures
and analyzed detailed pulsatile waveforms~\cite{Tate2026}. These results
motivate careful characterization of temporal sampling in human retinal
measurements.

Human applications now include analysis of blood-flow dynamics in glaucoma,
comparison with fluorescein angiography in infants with retinopathy of
prematurity, and retinal/choroidal flow indices in nonexudative age-related
macular degeneration~\cite{Vinnett2022,Shats2024,Linton2025}.
Recent work has further assessed intraday and interday repeatability of
optic-disc- and macula-centered human LSCI measurements
\cite{Wang2026Repeatability}. These studies motivate regional and temporal
endpoints while also requiring an explicit account of the vascular
territories and optical sampling underlying each index.

Here, we present a single-volunteer demonstration addressing sustained
computational throughput, retinal vascular visualization, and the evaluation
of rapid physiological dynamics. Human retinal images were acquired at
1000 frames/s and all incoming frames were processed with a pooled
$5\times5\times16$ contrast estimator in nonoverlapping temporal blocks.
This schedule yields 62.5 contrast maps/s, with rendering at 60 frames/s.
The nominal aggregation duration is 16 ms, compared with approximately
183 ms for the published 15-frame XyCAM analysis. This comparison describes
sampling parameters. The present demonstration reports the acquisition
and processing configuration, qualitative vascular visualization, and
visible cardiac-cycle variations in the principal retinal arteries and
veins in four peripapillary recordings from the same subject
(Supplementary Videos).

\begin{table}[!t]
 \caption{Representative retinal speckle-imaging implementations. Acquisition
 rates, exposure durations and statistical aggregation are separate quantities.
 Entries describe published configurations, not universal device limits.
 The table is not a claim of exhaustive priority or matched performance.}
 \label{tab:literature}
 \centering
 \small
 \setlength{\tabcolsep}{3pt}
 \renewcommand{\arraystretch}{1.12}
 \begin{tabular}{@{}p{0.30\linewidth}p{0.64\linewidth}@{}}
 \hline
 {\raggedright Implementation\par} & {\raggedright Acquisition and analysis\par} \\
 \hline
 {\raggedright Human LSFG-NAVI~\cite{Luft2016}\par} & {\raggedright 30 frames/s; 2-ms exposure. 118 images over about 4 s; cardiac-cycle processing.\par} \\
 {\raggedright Human XyCAM RI~\cite{Cho2020}\par} & {\raggedright 82 frames/s; camera exposure not specified in the cited study. $5\times5\times15$ kernel; nominal support about 183 ms.\par} \\
 {\raggedright Human retinal MESI~\cite{Qiu2022}\par} & {\raggedright Multi-exposure sequence: 0.5--64 ms acquired. 104 images over about 4.7 s; 2--64 ms retained for a fitted flow map.\par} \\
 {\raggedright Murine retinal LSCI~\cite{Tate2026}\par} & {\raggedright 376.2 frames/s; 2.5-ms exposure. $5\times5$ spatial contrast followed by signal filtering.\par} \\
 {\raggedright Present human recordings\par} & {\raggedright $3840\times2160$ at 1000 frames/s; $988\,\mu\mathrm{s}$ exposure. $5\times5\times16$ pooled contrast; nominal support 16 ms; 62.5 contrast maps/s; 60-frame/s rendering.\par} \\
 \hline
 \end{tabular}
\end{table}

\section{Materials and methods}

\subsection{Participants and experimental protocol}

Four recordings from the right eye of the same healthy 25-year-old
emmetropic man are reported. Images were acquired after written informed
consent. The procedures adhered to
the Declaration of Helsinki. Imaging was performed without pharmacological
pupil dilation in a dark room, using a standard head and chin rest and no
fixation target. Recordings were separated by a few seconds to allow
blinking. Imaging targeted the peripapillary retina in the neighborhood
of the optic disc. Each recording comprised 4096 successive frames at a
1-ms interval, corresponding to a nominal duration of 4.10 s. All 16384
frames in the four reported recordings were retained. No physiological
reference signal was recorded.

\subsection{Optical acquisition and detector sampling}

Illumination was provided by a volume-holographic-grating laser diode
(Thorlabs FPV852P) at $\lambda=852$ nm. In the illumination arm, an
engineered diffuser (Thorlabs
\href{https://www.thorlabs.com/item/ED1-C20-MD}{ED1-C20-MD};
25.4-mm diameter, $20^\circ$ circular top-hat profile, SM1-threaded mount)
was followed by a mounted N-BK7 biconvex lens (Thorlabs
\href{https://www.thorlabs.com/item/LB1811-B-ML}{LB1811-B-ML};
25.4-mm diameter, 35-mm focal length, antireflection coating for
650--1050~nm) to collimate the diffuse beam. A biconvex eyepiece (Thorlabs
\href{https://www.thorlabs.com/item/LB1106-B}{LB1106-B};
50.8-mm diameter, 125-mm focal length) provided diffuse Maxwellian
illumination and collected the fundus return, following the
diffuse-illumination approach of Bratasz and colleagues
\cite{Bratasz2022Diffuse}. Cross-polarized light backscattered by the
fundus was collected through the eyepiece, selected with a laser-line
polarizing cube beamsplitter (PBS; Edmund Optics, stock no.~\href{https://www.edmundoptics.com/p/35mm-850nm-laser-line-polarizing-cube-beamsplitter/7649/}{47-570};
35-mm cube, 850-nm design wavelength), and recorded by a Phantom S980 camera (Ametek Vision Research) with a
$5\,\mu\mathrm{m}$ sensor pixel pitch. The sensor was
placed at the focal plane of the eyepiece, with the optical system
conjugating the retinal and sensor planes. The cornea was positioned
125 mm in front of the eyepiece. No reference beam was used; the camera
directly recorded the fundus intensity sequence.
Laser linewidth, pupil and effective-aperture diameters, and speckle size
were not independently characterized. The retinal sampling and field below
are nominal geometric estimates.

Retinal intensity sequences were recorded at $\fs=1000$ frames/s
($\Delta t=1$ ms) over $W\times H=3840\times2160$ pixels with an
8-bit intensity depth. The exposure was $\Te=988\,\mu\mathrm{s}$,
giving a nominal interframe gap of $12\,\mu\mathrm{s}$ and a 98.8\%
integration duty cycle. Two Euresys Coaxlink QSFP+ frame grabbers
transferred the data to CPU RAM in an Intel Xeon workstation, then to
an NVIDIA RTX 5090 GPU. All incoming frames were processed, corresponding
to 8.29 Gpixel/s, or an uncompressed 8-bit payload of 8.29 GB/s
(66.35 Gbit/s), excluding transfer overhead.
Raw intensity recordings were saved at 8-bit depth; contrast maps were
displayed in real time.

Assuming 17 mm from the posterior nodal point to the retina in the
emmetropic eye, the estimated retinal pitch is
$p_{\mathrm r}=(5\,\mu\mathrm{m})(17/125)\simeq0.68\,\mu\mathrm{m}$
per pixel~\cite{Hastings2022Magnification}. This gives approximately
$2.61\times1.47$ mm full-frame coverage
($8.78^\circ\times4.95^\circ$) and a
$3.4\times3.4\,\mu\mathrm{m}$ footprint for the $5\times5$ kernel.
These geometry-based estimates assume focal-plane sensor placement and
no additional magnification. Individual calibration and usable-field
assessment were not performed. Pixel pitch does not specify optical resolution.

\begin{figure}[ht]
 \centering
 \includegraphics[width=\linewidth]{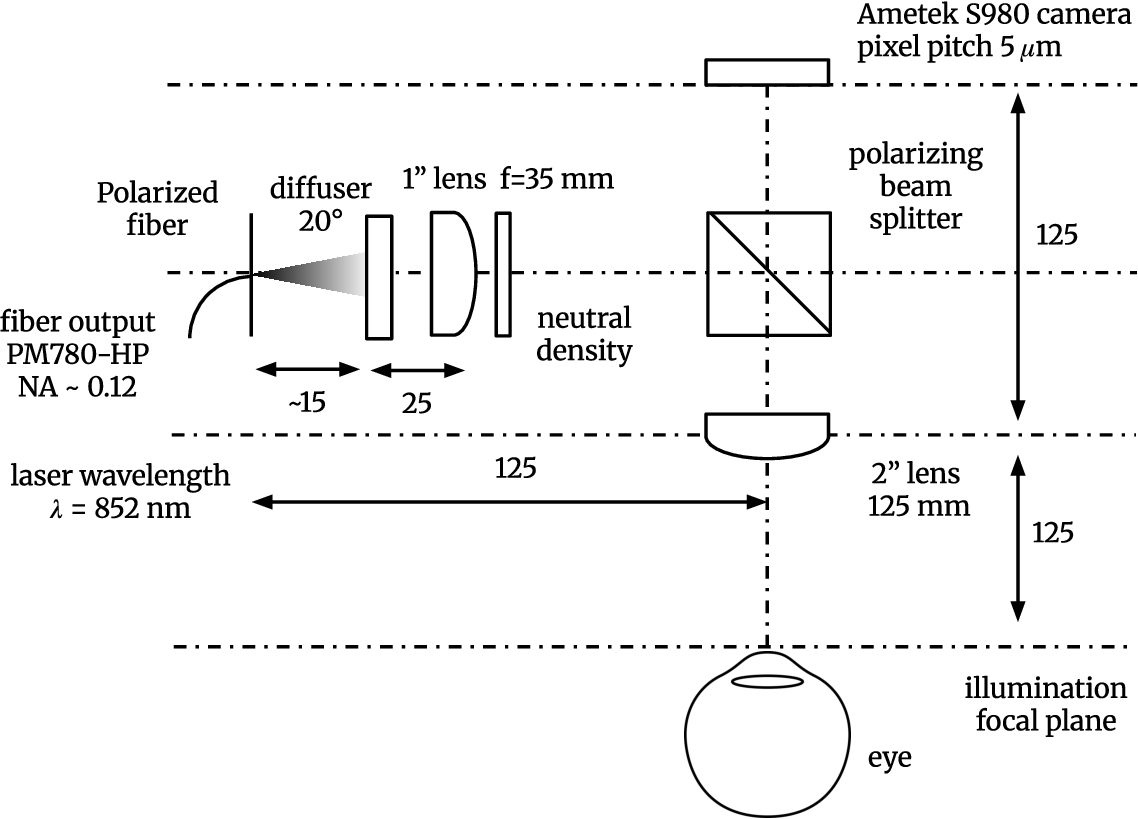}
 \caption{Acquisition setup for high-speed laser speckle contrast imaging
 of the human retina. Fiber-delivered laser light at $\lambda=852$~nm
 (PM780-HP fiber, numerical aperture $\simeq0.12$) passes through a
 $20^\circ$ engineered diffuser (Thorlabs ED1-C20-MD), a collimating lens
 (Thorlabs LB1811-B-ML, $f=35$~mm), and a neutral-density filter.
 A polarizing beamsplitter (PBS; Edmund Optics 47-570) directs the
 illumination toward a shared illumination and collection eyepiece
 (Thorlabs LB1106-B, $f=125$~mm).
 The cornea is positioned 125~mm from this lens for diffuse
 Maxwellian illumination. Cross-polarized fundus backscatter is collected
 through the same lens and recorded directly by the Phantom S980 camera
 ($5\,\mu\mathrm{m}$ sensor pixel pitch), without a reference beam.
 Longitudinal distances are indicated in millimetres.}
 \label{fig:setup}
\end{figure}

\subsection{Illumination and ocular exposure}

The laser was continuously on, with a nominal optical power of 5.0 mW
entering the eye at 852 nm. Diffuse illumination followed the
approach described by Bratasz and colleagues~\cite{Bratasz2022Diffuse},
with the reported eyepiece of 125 mm focal length.
Camera exposure and illumination duration are separate quantities: the
$988\,\mu\mathrm{s}$ integration and 98.8\% detector duty cycle do
not gate the continuous illumination. The four reported recordings represent
16.38 s of aggregate acquisition time, excluding the intervening
illumination and alignment periods. Total illumination duration was about
2 min.

\subsection{Speckle contrast processing and timing}

Speckle contrast was computed as $K=\sigma/\mu$, where $\mu$ and
$\sigma$ are the mean and population standard deviation of intensity
within a $5\times5$-pixel neighborhood spanning 16 frames
\cite{Qiu2010}. All incoming frames were processed in nonoverlapping
16-frame blocks, giving 400 samples per interior neighborhood and one
contrast map every 16 ms ($f_{\mathrm{map}}=62.5$ maps/s). Each
4096-frame recording contained 256 complete blocks. The exposure support
of a block was 15.99 ms. Maps were displayed at 60 frames/s with no
additional averaging.

Processing used the direct CUDA kernel backend of Holoflow, branch
\texttt{feat/contrast-pipeline}, commit \texttt{3d87195eb124}
\cite{Holoflow2026}. The implementation uses 32-bit floating point,
clips spatial neighborhoods at image borders, returns zero for
non-positive local means, and clamps negative computed variances to zero.
Block cadence and exposure support describe acquisition timing; processing
and acquisition-to-display latency were not independently benchmarked.

\begin{table}[!t]
 \caption{Experimental configuration and derived sampling quantities.
 Pixel rate and temporal support are distinct from measured latency and
 physiological bandwidth. Retinal scale and field are nominal geometry
 estimates without individual calibration.}
 \label{tab:parameters}
 \centering
 \footnotesize
 \setlength{\tabcolsep}{3pt}
 \renewcommand{\arraystretch}{1.08}
 \begin{tabular}{@{}p{0.46\linewidth}p{0.48\linewidth}@{}}
 \hline
 {\raggedright Quantity\par} & {\raggedright Value\par} \\
 \hline
 {\raggedright Participant\par} & {\raggedright Healthy man, age 25; right eye\par} \\
 {\raggedright Pupil / imaged region\par} & {\raggedright Undilated; peripapillary retina around the optic disc\par} \\
 {\raggedright Wavelength / entering power\par} & {\raggedright 852 nm / 5.0 mW\par} \\
 {\raggedright Camera / input format\par} & {\raggedright Phantom S980; $3840\times2160$\par} \\
 {\raggedright Sensor pixel pitch\par} & {\raggedright $5\,\mu\mathrm{m}$\par} \\
 {\raggedright Acquisition rate / interval\par} & {\raggedright 1000 frames/s / 1 ms\par} \\
 {\raggedright Exposure / integration duty\par} & {\raggedright $988\,\mu\mathrm{s}$ / 98.8\%\par} \\
 {\raggedright Recordings\par} & {\raggedright 4 reported; 4096 frames (4.10 s) each\par} \\
 {\raggedright Raw-input rate\par} & {\raggedright 8.29 Gpixel/s\par} \\
 {\raggedright GPU / host CPU\par} & {\raggedright NVIDIA RTX 5090 / Intel Xeon\par} \\
 {\raggedright Frame grabbers\par} & {\raggedright Two Euresys Coaxlink QSFP+\par} \\
 {\raggedright Contrast kernel\par} & {\raggedright $5\times5\times16$; 400 samples\par} \\
 {\raggedright Nominal / physical support\par} & {\raggedright 16 ms / 15.99 ms\par} \\
 {\raggedright Temporal stride\par} & {\raggedright 16 frames; nonoverlapping blocks\par} \\
 {\raggedright Contrast-map production\par} & {\raggedright 62.5 maps/s; 256 blocks/record\par} \\
 {\raggedright Rendering / extra averaging\par} & {\raggedright 60 frames/s / none\par} \\
 {\raggedright Preprocessing / observable\par} & {\raggedright None / $K=\sigma/\mu$\par} \\
 {\raggedright Recorded bit depth\par} & {\raggedright 8 bits/pixel\par} \\
 {\raggedright Nominal retinal sampling\par} & {\raggedright $0.68\,\mu\mathrm{m}$/pixel (estimated)\par} \\
 {\raggedright Nominal retinal field\par} & {\raggedright $2.61\times1.47$ mm (estimated)\par} \\
 {\raggedright Sensor angular field\par} & {\raggedright $8.78^\circ\times4.95^\circ$ (estimated)\par} \\
 \hline
 \end{tabular}
\end{table}

\subsection{Data retention and regional evaluation}

No preprocessing was applied: there was no dark-offset subtraction,
flat-field or bad-pixel correction, saturation masking, illumination
normalization, noise-variance subtraction, motion correction, or blink
rejection. All recorded frames were used. The signal quality was judged
visually suitable for vascular observation, and no saccade-associated
flashes were noted during real-time viewing. Signal-to-noise ratio,
saturation fraction, and detector noise were not quantified.

Evaluation concerned retinal vessel visibility and cardiac-cycle
variations in the principal retinal arteries and veins in the live $K$
maps and Supplementary Videos. Regional traces were obtained by spatially
averaging $K$ within an arterial, a venous, and an intervessel region, then taking
$1/\bar K$ in arbitrary units. Regional traces contained
255 samples per region, 16 ms apart, without smoothing or further
temporal averaging. Nominal block-start times span 16--4080 ms.
No statistical comparison or absolute flow
calibration was performed. There was no ECG, pulse, or independent
flow reference, kernel-length comparison, or repeatability analysis.
All four reported recordings were obtained from the same subject and eye.

\section{Results}

\subsection{Acquisition and analysis operating point}

The operating point combined $3840\times2160$-pixel, 8-bit acquisition
at 1000 frames/s and $988\,\mu\mathrm{s}$ exposure with online
processing on an NVIDIA RTX 5090. Four reported recordings, each containing
4096 frames, were processed with a $5\times5\times16$ kernel and
nonoverlapping temporal blocks. The resulting production cadence was
62.5 maps/s, with 60-frame/s rendering and no additional averaging
(Table~\ref{tab:parameters}). All 16384 frames in these recordings were retained.
The nominal raw-input rate was 8.29 Gpixel/s, corresponding to a
66.35-Gbit/s 8-bit payload before transfer overhead.

The camera interval was 1 ms, the nominal analysis window 16 ms, and its
exposure support 15.99 ms. Relative to the 15-frame XyCAM analysis at
82 frames/s, the nominal aggregation interval was shorter by
$(15/82)/(16/1000)\simeq11.4$. This parameter comparison does not
establish a matching improvement in physiological bandwidth, precision,
or diagnostic performance. Frame-loss counters, queue occupancy, and
processing and display latency were not measured, and longer uninterrupted
runs were not evaluated.

\subsection{In vivo retinal contrast maps}

\begin{figure}[!htbp]
 \centering
 \href{https://doi.org/10.5281/zenodo.23540440}{\includegraphics[width=\linewidth]{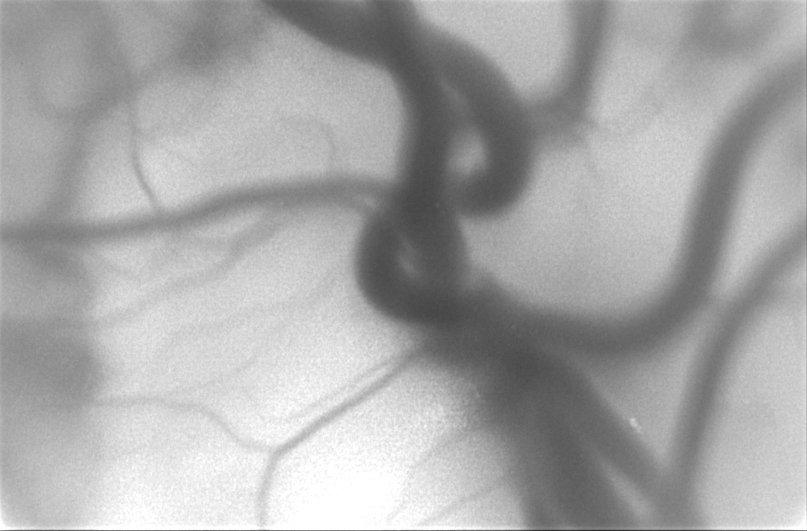}}
 \caption{Retinal vascular visualization by laser speckle contrast imaging.
 Representative contrast map $K=\sigma/\mu$ in the peripapillary retina
 near the optic disc, computed over $5\times5$ spatial pixels and
 16 successive frames acquired at 1000 frames/s with a
 $988\,\mu\mathrm{s}$ exposure. Principal retinal vessels and smaller
 branches appear dark against the surrounding fundus in this grayscale
 display. The \href{https://doi.org/10.5281/zenodo.23540440}{corresponding supplementary video}
 combines four recordings from the same subject and illustrates arterial
 and venous variations during the cardiac cycle.}
 \label{fig:maps}
\end{figure}

Principal retinal arterial and venous branches were visible in the
peripapillary region around the optic disc, together with
smaller retinal vessels. These observations were qualitative and based
on real-time contrast viewing. No vessel/background separation metric
or automated anatomical classification was calculated.
Figure~\ref{fig:maps} shows a representative speckle-contrast image.

\subsection{Arterial and venous variations during the cardiac cycle}

\begin{figure}[!htbp]
 \centering
 \includegraphics[width=\linewidth]{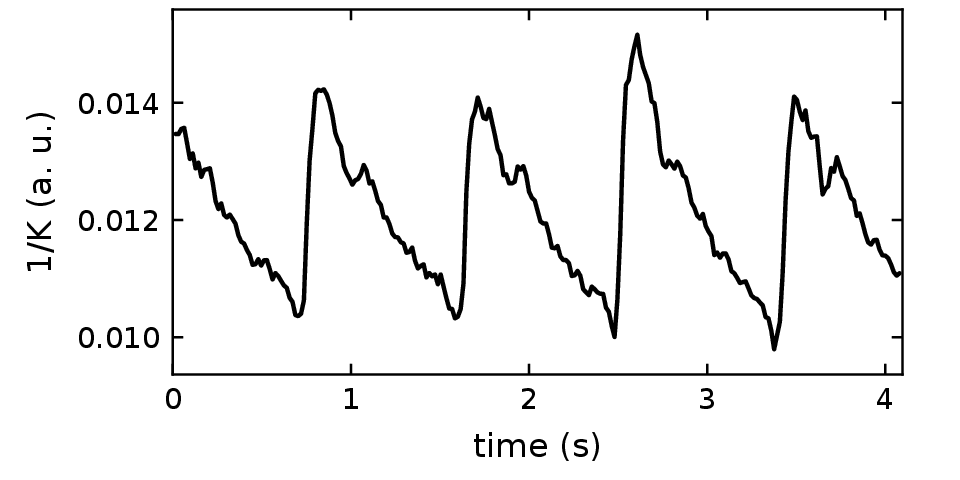}\par\smallskip
 \includegraphics[width=\linewidth]{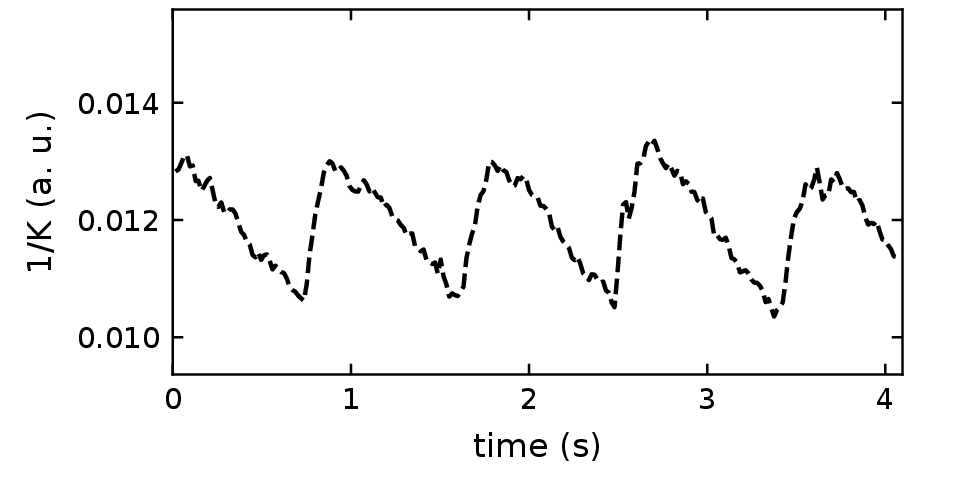}\par\smallskip
 \includegraphics[width=\linewidth]{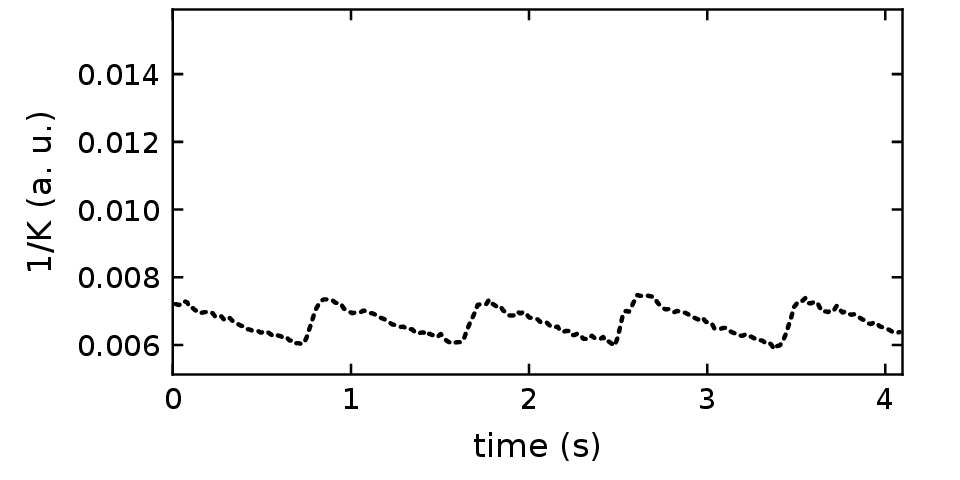}
 \caption{Arterial, venous, and intervessel signals in the peripapillary retina.
 Reciprocal of the spatially averaged speckle contrast, $1/\bar K$,
 in an artery (top, solid line), a vein (middle, dashed line), and
 an intervessel region (bottom, dotted line), in arbitrary units.
 All panels use the same axis scales.
 Samples are 16 ms apart (16 frames at 1000 frames/s).
 The arterial dicrotic notch and spontaneous venous pulsation are
 also visible in the
 \href{https://doi.org/10.5281/zenodo.23540440}{supplementary video}.}
 \label{fig:pulsatility}
\end{figure}

Variations in blood-flow-related contrast during the cardiac cycle were
visible in the principal retinal arteries and veins in the neighborhood
of the optic disc (Supplementary Videos). The videos illustrate these
variations in the four reported recordings from the same subject.
Figure~\ref{fig:pulsatility} shows regional arterial, venous, and intervessel
inverse-contrast traces. The arterial trace exhibits a dicrotic notch,
and the venous trace exhibits spontaneous pulsation. These pulsations
are also visible in the \href{https://doi.org/10.5281/zenodo.23540440}{supplementary video}.
Pulse amplitudes, inter-vessel timing differences, and independently
synchronized cardiac phases were not quantified.

\subsection{Data path and computational throughput}

The camera stream passed through two Euresys Coaxlink QSFP+ frame grabbers
to CPU RAM and then to the RTX 5090. The direct kernel backend computed
contrast from all incoming frames, while raw recordings were saved at
8-bit depth and contrast maps were displayed online. The stated input and
map rates describe the operating configuration; there is no separate
latency distribution, queue-depth trace, or independent frame-loss
benchmark for these recordings.

\section{Discussion}

\subsection{Position relative to retinal speckle imaging}

The present operating point extends the camera sampling rate used in the
human retinal implementations summarized in Table~\ref{tab:literature}.
The estimator belongs to established spatiotemporal LSCI
\cite{Qiu2010}. The comparison with the XyCAM implementation is particularly
direct in its spatial support, because both use a $5\times5$ neighborhood,
while its 15-frame temporal support differs slightly from the present
16-frame kernel~\cite{Cho2020}. The higher camera sampling rate shortens
the duration represented by this temporal aggregation. It does not by
itself establish improved contrast precision or clinical utility.
The reported implementation processes all input frames at the 8.29-Gpixel/s
operating point and generates contrast maps at 62.5 maps/s. Its biological
scope is one healthy eye across four reported peripapillary recordings.
Vascular visibility and cardiac-cycle variations in principal retinal
arteries and veins were observed in the Supplementary Videos and
regional inverse-contrast traces, while physiological
reference validation and independent timing benchmarks were not performed.

\subsection{What higher camera sampling contributes}

A shorter aggregation interval could preserve faster changes of a regional
contrast signal and reduce mixing of different phases of a cardiac waveform.
Whether it does so with sufficient precision depends on the measured
signal-to-noise ratio and temporal response. The benefit must be evaluated
against photon statistics, detector noise, eye movement and the number of
effectively independent speckles. The 400 samples in each kernel are a
nominal count; increasing camera sampling can also increase interframe
correlation at a fixed exposure and underlying decorrelation time.
The spatial neighborhood likewise spans a physical retinal area determined
by magnification, which has to be reported before spatial resolution can be
assessed~\cite{Kirkpatrick2008,Thompson2011}.

Intensity digitization is also relevant to the present 8-bit recordings.
Quantization can produce an intensity-dependent contrast bias
\cite{Song2013Quantization}, while alternative numerical implementations
of spatial contrast can change subsequent MESI estimates
\cite{Chammas2025}. A future quantitative pipeline should therefore be
validated against a direct high-precision calculation with matching
variance, window, and border conventions, and calibrated for the measured
detector response.

High-speed acquisition should therefore be described through separate
quantities: exposure duration, frame interval, estimator support, analysis
stride, additional filtering, and measured processing latency when relevant.
The acquisition rate alone does not establish real-time processing.
Processing is performed online on an RTX 5090, producing one map per
nonoverlapping 16-frame block (62.5 maps/s), with rendering at 60 frames/s
and no additional averaging. Four approximately 4.1-s recordings are
reported. Longer-run stability, frame-loss behavior, queue occupancy,
and processing/display latency were not independently characterized.

\subsection{Real-time processing and throughput benchmarks}

Comparisons must distinguish camera input, contrast conversion, and
flow-model inference (Table~\ref{tab:throughput}). Denoising
timings~\cite{Zhang2026Wavelet} and sensor ceilings or burst
captures~\cite{Dragojevic2015,Sun2015FPGA} describe further distinct
stages. The listed pixel rates normalize published image sizes and
timings; they do not provide a matched comparison of hardware or estimators.

\begin{table}[!t]
 \caption{Selected processing and multi-exposure benchmarks.
 Raw-input rates, processing-stage capacities, and fitted-map output
 are distinguished. Pixel rates are calculated from reported image
 dimensions and timings.}
 \label{tab:throughput}
 \centering
 \small
 \setlength{\tabcolsep}{3pt}
 \renewcommand{\arraystretch}{1.08}
 \begin{tabular}{@{}p{0.30\linewidth}p{0.64\linewidth}@{}}
 \hline
 {\raggedright Implementation\par} & {\raggedright Configuration and scope\par} \\
 \hline
 {\raggedright Laparoscopic GPU~\cite{Zheng2018Laparoscopic}\par} &
 {\raggedright $2048\times2048$ at 89 frames/s; 0.37 Gpixel/s processing.\par} \\
 {\raggedright Smartphone~\cite{Wu2023Smartphone}\par} &
 {\raggedright $4000\times3000$ at 20 Hz; 0.24 Gpixel/s processing.\par} \\
 {\raggedright Neurosurgical CPU~\cite{Selvaraj2026DASHING}\par} &
 {\raggedright $1280\times1024$ contrast conversion in 0.33 ms; 3.97 Gpixel/s stage capacity. Acquisition: 60 frames/s.\par} \\
 {\raggedright FPGA/SoC MESI~\cite{Hultman2020Realtime,Hultman2023Speed}\par} &
 {\raggedright $1280\times1024$ input at 1000 frames/s; 1.31 Gpixel/s. Continuous $320\times256$ perfusion maps at 15.6 frames/s.\par} \\
 {\raggedright REMI~\cite{Rivera2023REMI}\par} &
 {\raggedright $512\times512\times15$ contrast cube in about 0.12 s; approximately 8 inferred maps/s.\par} \\
 {\raggedright Neural MESI~\cite{Yu2023MESI}\par} &
 {\raggedright $700\times700$ map in about 400 ms; approximately 2.5 inferred maps/s.\par} \\
 {\raggedright Present operating point\par} &
 {\raggedright 8.29-Gpixel/s raw input; $3840\times2160$ contrast maps at 62.5 maps/s; 60-frame/s rendering.\par} \\
 \hline
 \end{tabular}
\end{table}

The present 8.29-Gpixel/s input, 62.5-map/s contrast production, and
60-frame/s rendering describe separate stages of four approximately
4.1-s recordings. Independent processing-latency, frame-loss, and
longer-run stability measurements remain necessary to characterize
sustained performance.

\noindent\textbf{Prospects for real-time multi-exposure imaging.}
A future MESI pipeline could sum $m=1,2,4,8,16,32,64$ successive raw-intensity
frames and recompute spatial contrast at nominal 1--64-ms synthetic
exposures~\cite{Hultman2018,Chammas2022Synthetic}. This exposure-dependent
measurement would extend the present pooled estimator toward
decorrelation-time estimation~\cite{Parthasarathy2008,Qiu2022}.
Candidate accelerators include a calibrated reduced-parameter GPU lookup
table, REMI or spatial-REMI~\cite{Rivera2023REMI,Rivera2023Software},
and neural inference~\cite{Hultman2020Realtime,Hultman2023Speed,Yu2023MESI}.
Exposure-subset optimization~\cite{Kazmi2014Optimization} could reduce
processing, while programmable multi-tap CMOS
acquisition~\cite{Sivakumar2019MultiTap} provides a sensor-level alternative.
This seven-channel schedule requires a rolling buffer of at least
64 raw frames. Updating every 16 frames could target 62.5 inferred maps/s;
the 32- and 64-ms channels would overlap between updates and retain their
full temporal support. Channels below the native integration time require
shorter camera exposures.
The forward model must account for the $988\,\mu\mathrm{s}$ integrations,
$12\,\mu\mathrm{s}$ gaps, detector noise, and static-light contributions.
Parameter identifiability with correlated exposure channels, retinal
motion robustness, and agreement with independent flow measurements
require validation. This inferred-map rate remains a prospective target
for the retinal pipeline.

\subsection{Interpretation of contrast and physiological specificity}

The measured observable is a coefficient of variation of retinal intensity.
Its relationship to blood motion depends on exposure, speckle sampling,
static and dynamic scattering, and the selected ensemble
\cite{Duncan2008,Parthasarathy2008}. Local anatomical variation, illumination changes, and eye movement can
also affect the pooled contrast. Pulsatility alone does not calibrate
the signal in velocity or flow units.

Light-scattering models commonly assume exponential or Gaussian field
decorrelation, with detection factors accounting for coherence, sampling,
and static--dynamic field mixing
\cite{Bandyopadhyay2005,Zakharov2006,Duncan2008,Liu2024}.
Single- and multiple-scattering descriptions depend on particle motion
and photon paths; vessel geometry and sampling depth can modify the
recovered index~\cite{Davis2014,Kazmi2015}. These models provide
interpretive context; none was fitted to the present recordings.

LSFG-derived MBR, XyCAM blood-flow velocity indices, and RFV are numerical
observables in instrument-dependent or arbitrary units, rather than
interchangeable physical measurements
\cite{Sugiyama2014,Shiga2014,Cho2020}. RFV validation has included retinal
venous bifurcation conservation and comparisons with laser Doppler
velocimetry; a subsequent phantom and human Doppler-OCT study also
identified saturation at higher velocities
\cite{Shiga2014,Luft2016Validation}. These results support specified
relative measurements without establishing a universal conversion from
blur or contrast to absolute retinal flow. Likewise, fitting an inverse
correlation time does not remove dependence on the scattering model.
Cross-vessel comparison can remain sensitive to vessel diameter and the
number of dynamic scattering events~\cite{Kazmi2015,Liu2024}.

The flow-related contrast is a reduction of speckle visibility by motion,
not a direct measurement of blood volume, vascular density, or oxygen
saturation. Within the appropriate model and operating range, faster
decorrelation decreases $K$ and increases an inverse-contrast index.
Percentage changes and waveform features can be useful under stable
acquisition conditions, but nonlinear transformations can change waveform
shape; MBR, $1/K$, $1/K^2$, RFV and $1/\tau_c$ should each retain their own
definition. An absolute volumetric-flow estimate additionally requires
a validated velocity measurement and vessel cross-sectional geometry,
rather than assigning flow units to the contrast alone.

\subsection{Vascular territories sampled in the fundus}

Fundus speckle studies distinguish resolved retinal arterioles and venules,
optic nerve head vessels and tissue, regions between visible retinal
vessels, and identifiable larger choroidal vessels
\cite{Sugiyama2014,Calzetti2018,Lu2022}. In LSFG, mean MBR over the
whole optic disc, its vessel mask, and its tissue mask is often denoted
$\mathrm{MBR}_{\mathrm{A}}$, $\mathrm{MBR}_{\mathrm{V}}$, and
$\mathrm{MBR}_{\mathrm{T}}$, respectively. The tissue mask excludes
resolved large vessels and provides a capillary-associated optic nerve
head index; it does not resolve individual capillaries or their depth
\cite{Sugiyama2014}. Vessel RFV uses background correction to reduce
the underlying tissue contribution~\cite{Shiga2014}.

Outside the optic disc, a region without a resolved retinal vessel is not
necessarily a retinal-capillary-specific measurement. Correlations with
OCT-derived choroidal thickness and vessel volume indicate that the choroid
can dominate such LSFG signals~\cite{Lu2022}. Choroidal-vessel RFV and
its exercise response have been investigated independently
\cite{Calzetti2018}; recent AMD studies have also analyzed large-retinal-
vessel and choroid-dominated image territories separately
\cite{Linton2025}. These anatomical masks are not equivalent to axial
separation of the superficial, intermediate and deep retinal plexuses or
the choriocapillaris. Conventional full-field speckle maps have no intrinsic
OCT-like depth gate.

For the present recordings, regions should therefore be named by their
observed anatomy and segmentation, with ``capillary-associated'' or
``choroid-dominated'' used only where justified. Wavelength, pigmentation,
multiple scattering, focus, aperture, and motion can alter the relative
weighting of retinal and choroidal contributions. Baseline differences
between territories are consequently more difficult to interpret as
velocity differences than repeated changes within a consistently sampled
region~\cite{Sugiyama2014,Davis2014,Liu2024}.
The present 852-nm undilated recordings sample the peripapillary retina
around the optic disc. Cross-polarized collection does not itself isolate
retinal vascular layers, and the direct intensity estimator supplies no
axial separation. Without preprocessing or motion correction, the
territorial interpretation must also account for movement and local
intensity gradients. The nominal retinal scale requires individual
calibration, and speckle sampling remains to be characterized; capillary
resolution and layer-specific or absolute flow
are not established by the present configuration alone.

\subsection{Limitations and next validation steps}

The demonstration comprises four approximately 4.1-s recordings of the
peripapillary retina in the undilated right eye of one healthy 25-year-old
emmetropic man. Vessel visibility and cardiac-cycle variations were
assessed descriptively from images, videos, and regional inverse-contrast
traces. No statistical waveform analysis, reference flow measurements,
physiological synchronization, or numerical contrast-separation endpoints
were obtained. All recorded
frames were used, but dedicated acquisition-loss and latency tests were
not performed. The absence of visually obvious saccade-associated flashes
does not quantify motion sensitivity.

The geometric retinal scale was not individually calibrated. Pupil and
effective-aperture diameters, speckle sampling, detector noise, and signal
precision were not measured, and no motion or noise correction was applied.
An apparatus-specific exposure assessment under
ISO 15004-2:2024~\cite{ISO15004_2024} was not documented.
These limitations bound physiological and quantitative
interpretation of the present observations.

Repeatability and reference agreement require separate evaluation with
independent recordings and participant-level analysis. Published human
multi-exposure and commercial LSCI repeatability studies provide useful
comparators~\cite{Qiu2022,Wang2026Repeatability}, while reference agreement
addresses a separate question~\cite{Luft2016Validation,Shats2024}.
Clinical populations and physiological perturbations remain future work.
Retinal vasoreactivity and vasomotion have established LSCI precedents
\cite{Neganova2016,Neganova2017}; assessing slow regulation requires
recordings long enough to contain multiple cycles of the process of interest.

\section{Conclusion}

We demonstrate real-time human retinal LSCI using $3840\times2160$-pixel,
8-bit acquisition at 1000 frames/s and $988\,\mu\mathrm{s}$ exposure,
with online processing on an RTX 5090. Nonoverlapping $5\times5\times16$
contrast estimates produce 62.5 maps/s for 60-frame/s rendering, without
additional averaging. In four recordings from one undilated healthy right
eye, principal retinal arterial and venous branches and smaller vessels
were visualized in the neighborhood of the optic disc. Variations
during the cardiac cycle were visible in the principal retinal arteries
and veins (Supplementary Videos). Regional inverse-contrast traces
illustrate an arterial dicrotic notch and spontaneous venous pulsation.
These descriptive observations support high-throughput retinal vascular
visualization; quantitative physiological performance and independent
computational benchmarks remain to be established.

\begin{acknowledgments}
This work was supported by ANR LIDARO (ANR-22-CE19-0033-01) and IHU FOReSIGHT.
\end{acknowledgments}

\section*{Supplementary videos}
The \href{https://doi.org/10.5281/zenodo.23540440}{combined supplementary video} shows
the four recordings around the optic disc in the same subject, with
visible cardiac-cycle variations in flow-related contrast in the
principal retinal arteries and veins, including the arterial and
spontaneous venous pulsations illustrated in Fig.~\ref{fig:pulsatility}.

\section*{Data and code availability}
Raw 8-bit recordings will be made publicly available with this study;
repository details will accompany the data release. Processing used the
open-source Holoflow software, distributed under the Apache License 2.0,
available at \url{https://github.com/DigitalHolography/Holoflow}
\cite{Holoflow2026}. The experimental implementation was branch
\texttt{feat/contrast-pipeline}, commit \texttt{3d87195eb124}
(22 October 2026; version \texttt{0.2.0-dev.4}), using the direct CUDA
kernel backend. The bibliography pins the full commit. Raw images were
saved at 8-bit depth; $K$ was calculated in floating point and displayed
in real time.

\section*{Disclosures}
The authors declare no competing interests.

\bibliographystyle{apsrev4-2}
\bibliography{LSCI}

\end{document}